\def\devlaw{de Vaucouleurs' law}

\def\unit #1{\,{\rm #1}}
\def\jy{\unit{Jy}}
\def\kpc{\unit{kpc}}
\def\mhz{\unit{MHz}}
\def\msol{M_{\odot}}

\def\mean #1{\left< #1 \right>}
\def\onlyten#1{10^{#1}}
\def\chisqr{\chi^2_{\nu}}
\def\REFR{}

\def\bandr{B{\rm\ and \ }R}
\def\re{r_e}
\def\rer{r_e(R)}
\def\reb{r_e(B)}

\def\ie{i.\,e., }
\def\etal{et al.\,}
\def\eg{e.\,g.\,} 

\def\labsec #1{\label{sec:#1}}
\def\tab #1{Table~\ref{tab:#1}}
\def\secn #1{Section~\ref{sec:#1}}
\def\fig #1{Figure~\ref{#1}}
\def\eqn #1{Equation~\ref{eq:#1}}

\def\bcen{\begin{center}}
\def\ecen{\end{center}}
\def\bfig{\begin{figure}}
\def\efig{\end{figure}}
\documentstyle[11pt,aaspp4,epsfig]{article}

\slugcomment{To appear in ApJ Letters}

\lefthead{Mahabal et al.}
\righthead{Effective Radii and Color Gradients}

\begin{document}
\renewcommand{\textfraction}{0.02}
\title{Effective Radii  and Color Gradients in Radio Galaxies}
\author{Ashish Mahabal\altaffilmark{1} and Ajit Kembhavi}
\affil{Inter-University Centre for Astronomy and Astrophysics,
	Post Bag 4, Ganeshkhind, Pune~411~007, India}
\and
\author{P. J. McCarthy}
\affil{Observatories of the Carnegie Institute of Washington,
	Pasadena, CA, USA}
\altaffiltext{1}{present address: Physical Research Laboratory, A and A
	division, Navrangpura, Ahmedabad, India}
\begin{abstract}
We present de Vaucouleurs' effective radii in $\bandr$ bands 
for a sample of Molonglo Reference Catalogue radio galaxies and a control 
sample of normal galaxies. We use the ratio of the scale 
lengths in the two bands as an indicator to show that the radio 
galaxies tend to have excess of blue color in their inner region 
much more frequently than the control  galaxies. 
We show that the scale length ratio is a useful indicator of 
radial color variation
even when the conventional color gradient is too noisy to serve the purpose.

\end{abstract}
\section{Introduction}
\labsec{intro}

Traditionally, radio galaxies were believed to be ellipticals
consisting of a coeval population of old stars,
and almost no dust or gas. However, detailed photometric studies in
the optical band, as well as X-ray observations, have demonstrated that 
elliptical galaxies, especially those hosting radio sources,
not only have significant quantities of dust and gas, 
but also possess fine morphological structure indicating some 
amount of activity in
the past $\sim\onlyten{2}$ million years 
\REFR (see \eg\ Smith \& Heckman 1989).
Radio galaxies host an active galactic nucleus (AGN) and also 
have  radio jets which transport a very large amount of energy over hundreds
of kiloparsecs. Such phenomena are likely to be associated with morphological
features and star formation activity not found in normal elliptical galaxies.

In this paper we present the main results from a detailed morphological 
study of a sample of radio galaxies from the {\it Molonglo Reference Catalogue}
(MRC). We present de Vaucouleurs' effective radii (scale lengths), 
obtained from careful model fits to surface brightness profiles of 
the galaxies, and show that the ratio $\reb/\rer$, of the scale 
lengths in $\bandr$ filters, provides a measure of the color 
gradient in the galaxy.
The ratio is related to color gradients measured conventionally, but
is more robust: it can provide an estimate of the color gradient even
when the signal-to-noise ratio is not good enough for the color gradient
to be measured unambiguously using the conventional technique. Using
the ratio we show that a large fraction of radio galaxies become bluer
towards the center.

\section{Sample and observations}
Our results are based on the observations of
30 galaxies from the MRC which have $408\mhz$ radio 
flux   $S_{408} > 0.95\jy$, redshift  $z<0.3$ and declination
$-30\deg \le \delta(1950) \le -20\deg$. The objects were observed
from the Las Campanas
Observatory (LCO), Chile in Jan 1995
and Feb 1996 using the 1.0m f/7 Swope telescope
and the 2.5m f/7.5 Du Pont telescope.
Images were obtained in Johnson's $B$ and Cousin's $R$ filters, which are 
centered at $0.44\micron$ and $0.65\micron$ respectively and have a 
bandwidth of $\sim0.1\micron$.  The  typical exposure time was 
$\sim20\min$ in $R$ and $\sim60\min$ in $B$.
The FWHM of the PSF ranged from $\sim1''\!.1$ to $\sim1''\!.5$. When the
FWHM of the PSF was different for the images in $\bandr$ filters for the
same object, the better PSF was degraded to match the other before it was used
to compare properties that involve both the filters. The plate
scale was $0''.7/\unit{pixel}$ and the total field of view 
covered in an exposure was 
$11'.5 \times 11'.5$ in
Jan 1995 and $23' \times 23'$ in Feb 1996. 
The large field allowed us to determine the sky
background more accurately than is usually possible.
All the processing was done in the normal way using tasks from
IRAF and STSDAS, and the details will be published elsewhere 
\REFR (Mahabal, Kembhavi \& McCarthy 1999).

For the purpose of comparison we  extracted a control sample 
from the CCD fields of our radio galaxies.  The sample
consists of all non-radio, early type galaxies from the fields which 
have semi-major axis length $>\,15''$. There are 30 galaxies in the 
control sample, and these were  processed and analyzed in a fashion 
identical to the radio galaxies.
The redshifts for the galaxies in the control sample are
not known. However, the distribution of angular sizes and apparent
magnitudes for the control sample
are similar to that of the radio sample and hence the redshift distributions for
the two samples are unlikely to be too different.
Our main results, based on the ratio of scale lengths in
the $\bandr$ filters are unlikely to be affected by any small changes in the
redshift distribution.

\section{Surface photometry}
\labsec{surface_photometry}

We fitted the isophotes of
each galaxy in both $\bandr$ filters with a succession of ellipses
with different semi-major axis lengths  using tasks in {\it IRAF}
based on the algorithm described by \REFR Jedrzejewski (1987).
The mean surface brightness for the series of best-fit ellipses
gives us the radial surface brightness profile of a galaxy as a function of
semi-major axis length.  From the radial profile in the two bands we obtained
the $B-R$ color profile for each galaxy.    

Major contributions to the galaxian light can in general come from
a spheroidal bulge and a flattened disk.  In 
active galaxies an additional substantial contribution to
the central region can be made  by the active galactic nucleus (AGN)
acting as a point source. The relative strengths  of the  
components vary over galaxy type, 
and  can be determined by fitting the observed radial surface brightness profile
of the galaxy with a composite model made up of contributions from each
component.  

We have assumed that the bulge profile is described by 
de Vaucouleurs' law, and the disk profile by an exponential.  The contribution
of the AGN corresponds to a point source broadened by the point spread function.
We have found for our sample of radio galaxies that the AGN is too
weak to be unambiguously detected, and distinguished from other compact 
features which may be present.  We therefore do not include an AGN 
in our fits.  The model surface brightness $I(r)$ at semi-major axis length $r$
is then given by  
\begin{equation}
\label{eq:summed_profile}
I(r)=I_e\exp\{-7.67[(r/r_e)^{1/4}-1]\} + I_d \exp(-r/r_d)
\end{equation}
where $I_e$ is the bulge intensity at de Vaucouleurs' effective radius 
(scale length) $r_e$,  $I_d$ is the disk surface brightness at $r=0$, and $r_d$ 
is the disk scale-length. de Vaucouleurs' law is known to provide
a good fit over the range $0.1r_e<r<1.5r_e$  \REFR (Burkert 1993).  The
points that we use in our fits lie in this range.  

To obtain the best-fit model to an observed galaxy profile, 
we generate a  model galaxy with
two dimensional surface brightness distribution corresponding to trial
values of the four parameters in \eqn{summed_profile}, and observed bulge
and disk ellipticities.   We convolve
the model  with a Gaussian point-spread function (PSF) determined from the 
observed frames for the galaxy in question, and then obtain the model radial 
profile. We determine the best-fit parameters by minimizing the 
reduced chi-square
function $\chisqr=\sum(I_o-I_m)^2/\nu\sigma^2$,  where $I_0$ and $I_m$ are
the observed and model surface brightnesses at specific
distances along the semi-major axis.  The standard deviation $\sigma$ at each
point is obtained from the ellipse fitting task in IRAF and includes 
photon counting as well as ellipse fitting errors;   $\nu$ is the number of degrees of freedom, and is equal to the number of points used in the fit reduced
by the number of free parameters.  

Contributions to 
$\chisqr$ are obtained at semi-major axis lengths $r$ in the range $r_1<r<r_2$,
where the inner limit $r_1$ is chosen such that it lies at a radial separation 
of 1.5 times the full width at half maximum (FWHM) of the PSF and 
the outer limit $r_2$ is chosen where $\sigma(I)/I$ drops to 0.1.  We have
omitted points inside $r_1$, which typically involve just a few pixels,
from the fit  because the profile here can be seriously affected by the PSF, 
as well as by any departures 
from de Vaucouleurs' law that may be present close to the center.  The 
PSF influences the shape of the profile to several times the FWHM 
(see \REFR Franx, Illingworth, \& Heckman 1989;
\REFR Peletier \etal 1990), but we account for this in our work by
convolving the model profile with a model PSF before comparing it with the observed profile.  We have  carried out runs 
on artificial galaxies (created using the IRAF package {\it artgal}) and 
on nearby elliptical galaxies and galaxies from our sample which suggest 
that the value of the extracted effective radius stabilizes beyond 1.5 times the 
PSF FWHM \REFR (Mahabal 1998), when PSF convolved profiles are used. 

We have fitted the bulge plus disk combination in \eqn{summed_profile}
to the radio galaxies.  In many cases the 
ratio $D/B$ of the disk to bulge luminosity is $\ll1$, which is consistent 
with the radio galaxies being ellipticals.  In some cases we detect a disk component  $D/B\geq0.3$, but the disk scale length is small, except in two
cases, so that the 
detected ``disk'' is a small scale structure unlike the disks in 
spiral galaxies.  We will describe elsewhere our findings regarding the 
disk like structures, and devote our attention here to the bulges.  Pure
bulge fits also turn out to be acceptable in most cases, but the disk plus
bulge fits provide better $\chisqr$ values on the whole, and we retain them
as one of our aims in the larger investigation has been to find any disk like structures which may be present.  The results reported here would not change 
if pure bulge fits were used in the discussion.  None of the control galaxies 
has a significant disk component.

\subsection{Goodness of fit}

We get very good ($\chisqr<1$) or acceptable ($1<\chisqr<2$) fits 
in  $\sim85\%$ of the cases for the radio as well as the control sample.  
Visually too $<\sim10\%$ galaxies are seen to be highly distorted in
the radio sample. 
It follows that strong radio sources
{\it do not} prefer highly distorted galaxies.
$\chisqr$ does not increase as a function of redshift, \ie\
we get good fits right upto the redshift of 0.3 that we have
considered.
We find that $\chisqr(B)$ values are in general smaller than
the corresponding $\chisqr(R)$ values.  The lower values in $B$ are partly due
to the higher $\sigma$ values there.  It also appears that isophote distorting
influences like emission and absorption regions which could increase $\chisqr(B)$
are averaged out in the ellipse and profile fits.

\placefigure{fig1}
\bfig
\bcen
\mbox{\epsfig{figure=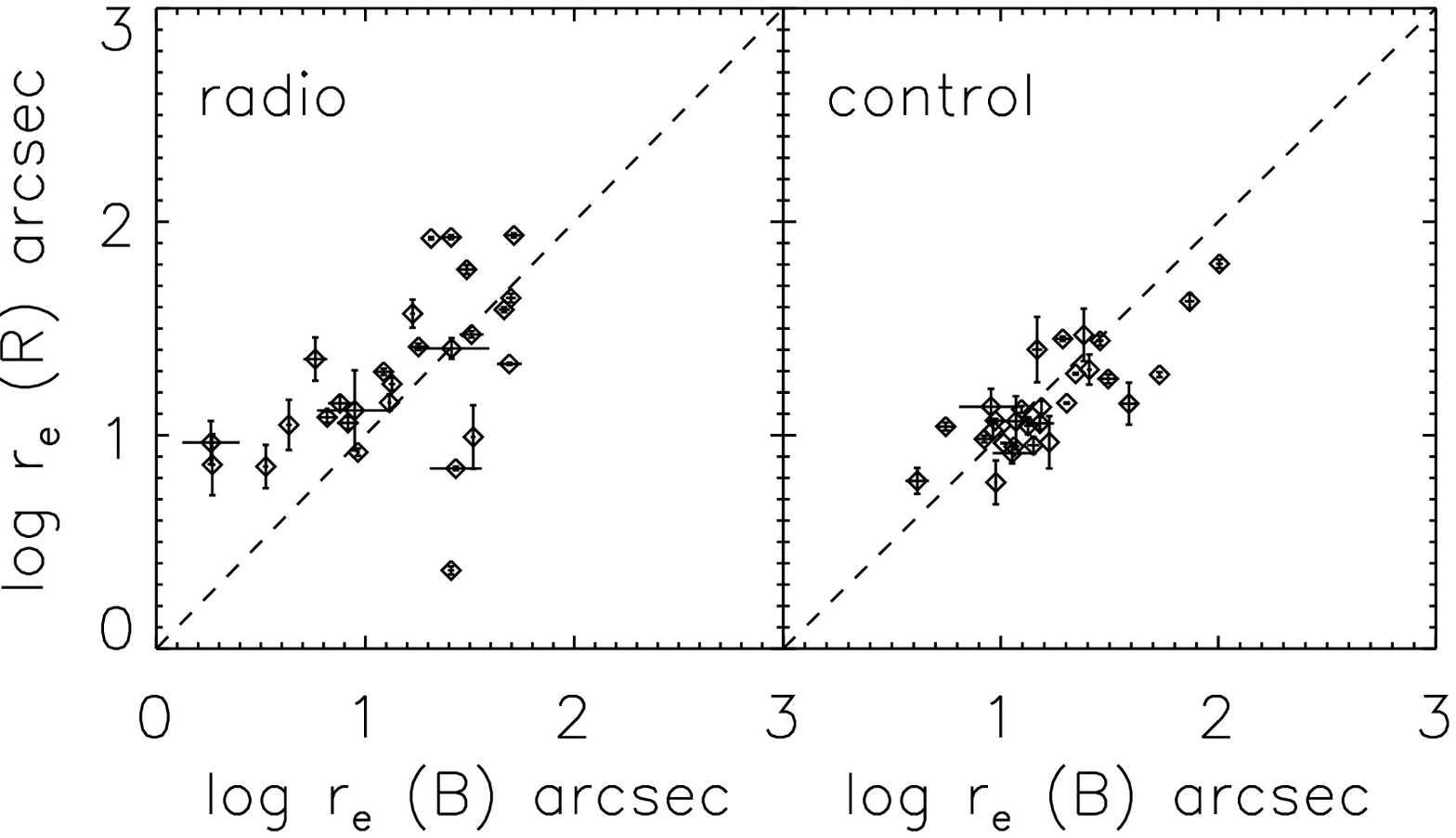,width=4in}}
\caption
{$\re$ in $\bandr$ filters for the radio sample (left)
and the control sample (right). The dotted line is the locus 
$\reb=\rer$.
Points above it denote galaxies which become bluer inwards. There is a
larger number of such cases amongst the radio galaxies.\label{fig1}
}
\ecen
\efig

\subsection{Bulge parameters}

We now turn to the bulge scale lengths.  \fig{fig1} shows a 
plot of $\reb$ against $\rer$ for the radio and control galaxies. The 
$1\sigma$ errors on $\re$, obtained from the fitting
program, are typically $\sim10\%$.  The scale lengths in the two filters
are equal to within $1\sigma$ in several cases, and these values are
scattered around the $\reb=\rer$ line in the figure.  In the other cases
we have $\reb<\rer$  (points above the equality line) or $\reb>\rer$ 
(points below). It is obvious from the figure that the former are
more numerous in the radio galaxies, while the latter occur more frequently 
in the control galaxies.  

When $\reb>\rer$, the surface brightness in $R$ increases more
rapidly towards the center than the surface brightness in B. 
In other words, from the definition of the effective radius $r_e$, 
half the red light from the galaxy is contained in a 
smaller region than half the blue light. $\reb>\rer$ therefore implies that
the galaxy on the average becomes redder inwards.
Similarly, when $\reb<\rer$, the galaxy becomes
bluer as one moves towards the center.  The distribution of points in
\fig{fig1} therefore show that the radio galaxies become bluer 
towards the center more often than the control galaxies.  

We have shown in \fig{fig2} the distribution of the ratio
$\reb/\rer$ for the radio and control galaxies.
An application of the Kolmogorov-Smirnov test shows that 
the two distributions are different at 99.99\% confidence level.
For the radio sample the mean value of the ratio is 
$\mean{\reb/\rer}_r=0.87 \pm 0.15$, while for the control sample the
mean value is $\mean{\reb/\rer}_c=1.25 \pm 0.10$.
The larger value of the ratio for the control sample is consistent with
previous studies of early type galaxies, which showed that their colors become
redder inwards  \REFR (\eg\ Sandage \& Vishwanathan 1978).
Contrary to this behavior, the distribution of the bulge scale length 
ratio for the radio galaxies shows that they tend to become bluer as one moves
towards their inner region, \ie\ the color variation in
radio galaxies is opposite of that in the control galaxies.
In the next section we will consider the relation between the 
scale length ratio and the conventional color gradient.  

\placefigure{fig2}
\bfig
\bcen
\mbox{\epsfig{figure=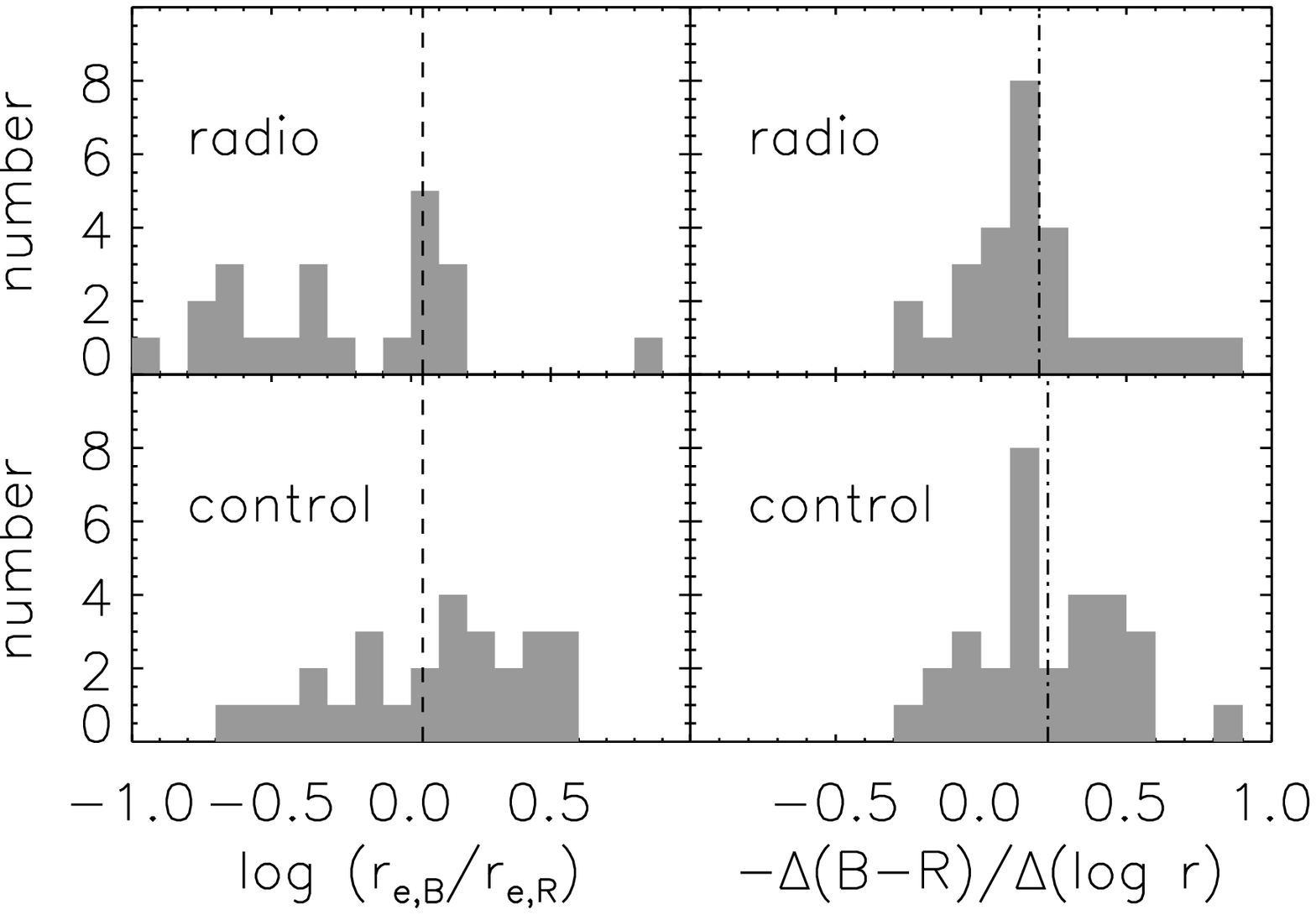,width=4in}}
\caption
{The distribution of $r_e(B)/r_e(R)$ (left) and the color gradient
(right) for the radio and control samples. Negative numbers on the
x-axis are indicative of galaxies with excess blue emission towards
the center.\label{fig2}}
\ecen
\efig

We have excluded from the profile fits points within 1.5 times the
FWHM of the PSF from the center.
For our sample the excluded region has a physical dimension extending to 
$\sim4\kpc$ at the highest redshift.   It follows that the inner bluer
color of the radio galaxies is not due to a blue AGN or other
unresolved features at the center, and must arise in regions which are 
spread out.  

\section{Color gradients}

The color variation in a galaxy is normally measured by a
color gradient parameter 
$G\equiv\Delta(B-R)/\Delta({\rm log\,r})$.
The change in color per decade in radius is almost linear in most
galaxies and $G$ 
is obtained by fitting a straight line to the color
profile between an inner radius $r_1$ and an outer radius $r_2$.
We choose these radii as described in \secn{surface_photometry}, with the
additional caveat that now $r_2=\min{(r_2(B),r_2(R))}$.

When two small galaxies (angular diameter $<15''$) and a quasar host
in our radio sample are excluded, we find that the mean color gradients, 
in magnitudes per $\unit{arcsec^2}$ per decade in radius, for 
the radio and control samples are $\mean{G}_r= -0.20\pm0.05$ and 
$\mean{G}_c= -0.23\pm0.05$. The distribution of gradients is shown in \fig{fig2}.
We find that the numbers that we obtain are larger in magnitude
than those obtained by other authors.
For a sample of normal ellipticals (with dusty galaxies excluded)
Peletier \etal\ (1990) \REFR had obtained
a color gradient of --0.1.
Zirbel \REFR (1996) had obtained a value of --0.15
for a sample of radio galaxies.
However, we have confirmed that the larger numbers we get are not
owing to photometric errors.

\subsection{Color gradients and scale length ratios}

Color gradients and scale length ratios are both indicative
of the change in color with distance from the center and they
are related, to a first approximation, by:
\begin{equation}
\label{eq:deltac2}
G\equiv\frac{\Delta(B-R)}{\log r_2/r_1}\simeq
\frac{2.06(r_2^{1/4}-r_1^{1/4})}{\reb^{1/4} \log (r_2/r_1)}
\left(1-\frac{\reb}{\rer}\right).
\end{equation}
For galaxies that obey \devlaw\ the color gradient can therefore 
be estimated  from the fitted bulge scale lengths.
We enumerate in \tab{colgrads} the distribution of color gradients 
obtained using the scale lengths, as well as the gradients obtained
directly from the color profiles using the measured values of the 
$B-R$ color at $r_1$ and $r_2$ 
We have deviated from the more usual custom of taking the outer point
at $r_e$ since in some cases the color
profile is very noisy at that radius.   The gradient does not
change very much with changes in $r_1$ and $r_2$.  It is seen from the table
that the color gradients obtained directly from the color profiles 
have a different distribution from the bulge scale length related
color gradient.  The distribution of the latter clearly shows that
radio galaxies become bluer towards 
the center while the control galaxies become redder.  This distinction is
not obvious from the distribution of the directly measured color gradient.   
In \fig{fig3} we have plotted the conventional color gradient against
that obtained from the scale lengths for the radio galaxies. 
A simple $2 \times 2$ contingency test 
shows that the two color indicators vary in the same sense.

\placetable{tab:colgrads}
\begin{table*}
\begin{center}
\begin{tabular}{lrrrr}
\tableline
\tableline
Color & \multicolumn{2}{c}{From color profile} &
		\multicolumn{2}{c}{From scale lengths}\\
gradient        & Radio & Control & Radio & Control \\
\tableline
$<0$            & 20 & 24 &  9 & 20 \\
$>0$            &  7 &  6 & 18 & 10 \\
uncertain       &  3 &  0 &  3 &  0 \\
\tableline
\end{tabular}
\end{center}
\tablenum{1}
\caption
{Color gradient details for the two samples as obtained
from the color profile and from the bulge
scale lengths. A negative color gradient is indicative
of a redder center relative to the outer regions.
\label{tab:colgrads}}
\end{table*}

\placefigure{fig3}
\bfig
\bcen
\mbox{\epsfig{figure=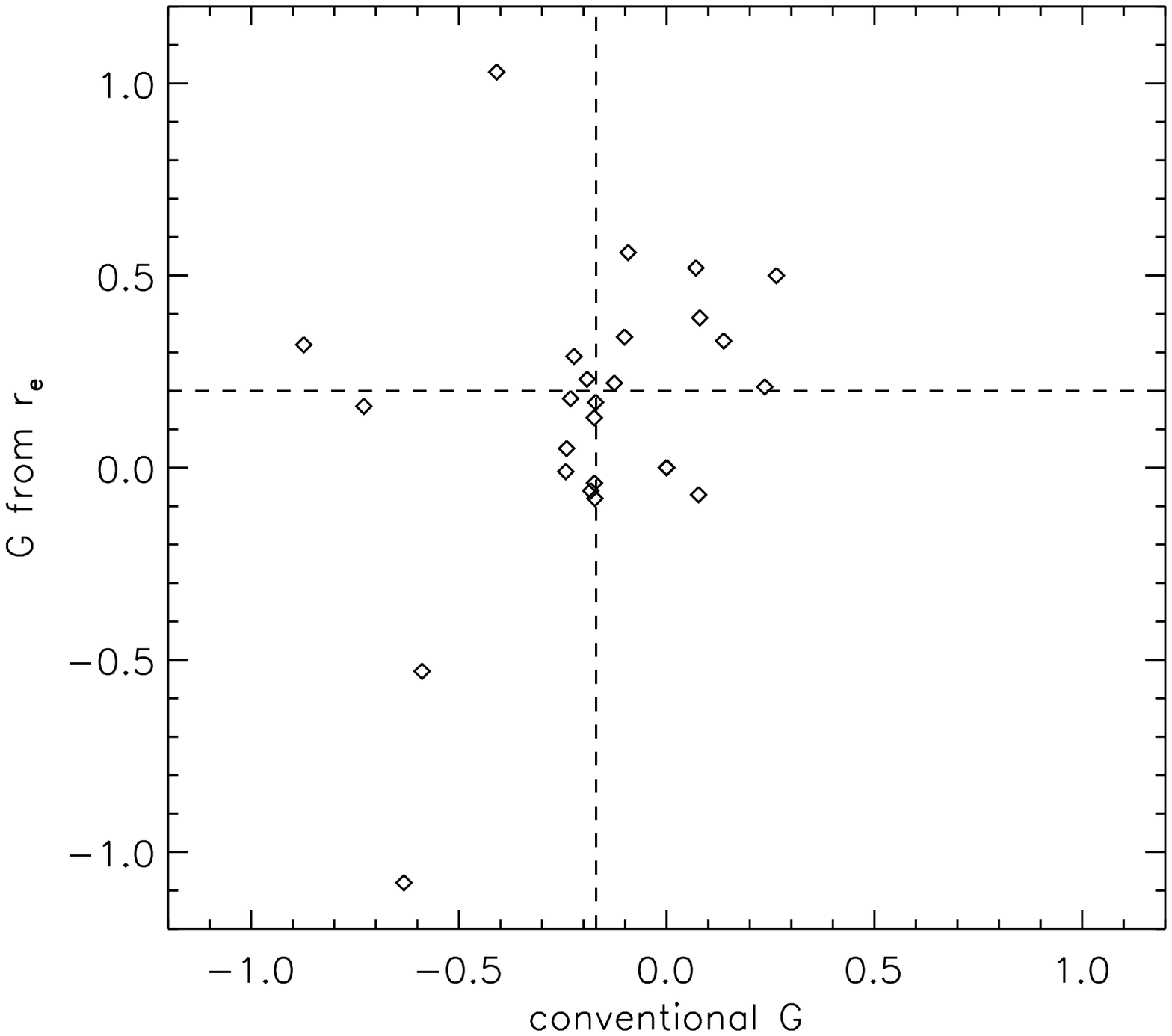,width=6in}}
\caption
{Conventional color gradient against color variation as given by
the $\bandr$ scale lengths for the radio galaxies. 
The latter is indicative
of bluer central regions in many cases. A $2 \times 2$ contingency test
shows that the two color indicators vary in the same sense.\label{fig3}}
\ecen
\efig

\placefigure{fig4}
\bfig
\bcen
\mbox{\epsfig{figure=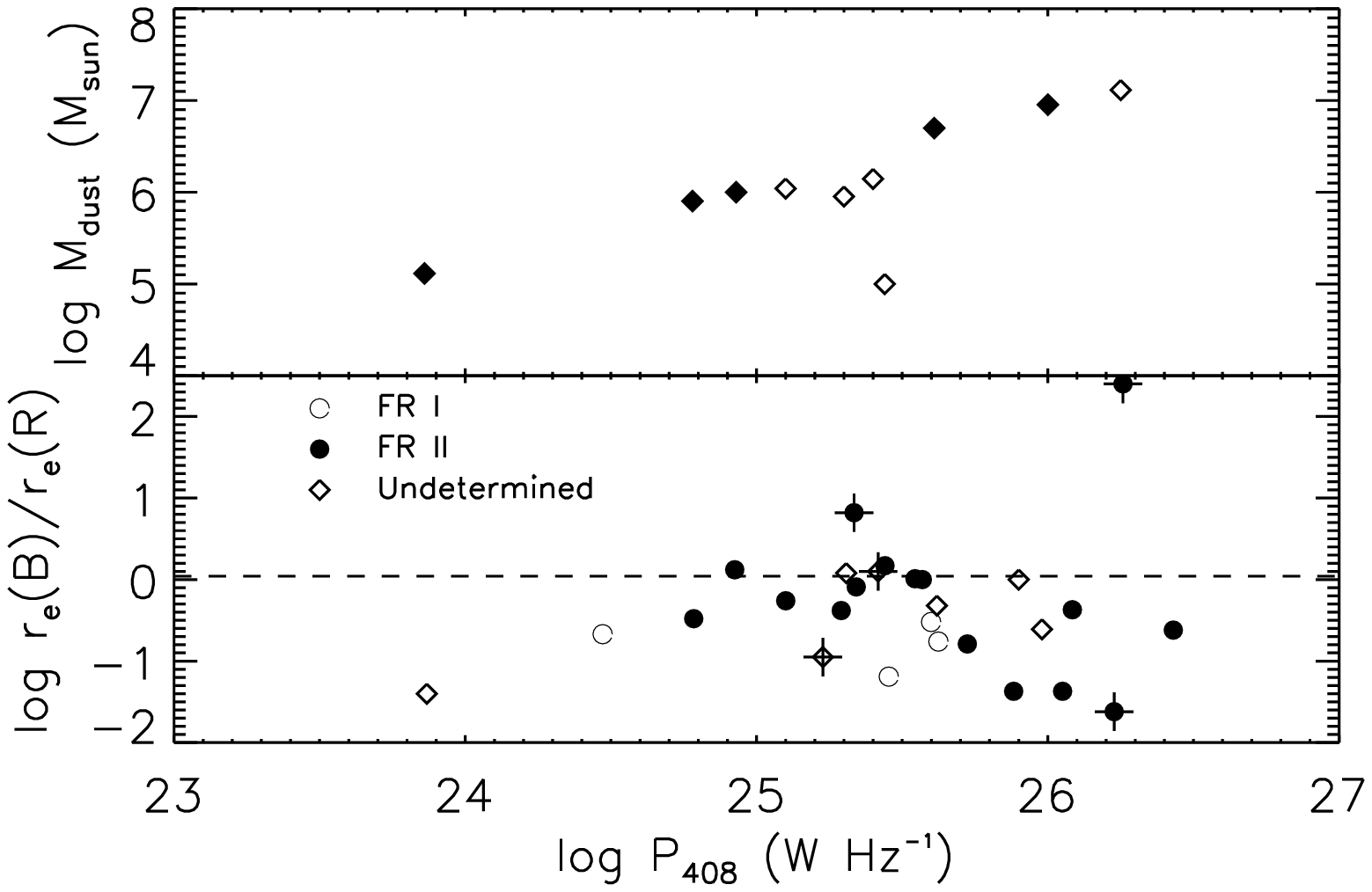,width=6in}}
\caption
{
Top: correlation of dust mass in radio galaxies with dust lanes
(filled diamonds) or dust patches (open diamonds)
at the center with radio power.
Bottom: $\reb/\rer$ as a function of radio
power. The dashed line shows the scale length ratio expected for
a normal galaxy. There is a hint that more powerful galaxies are bluer
towards the center.
\label{fig4}}
\ecen
\efig

The conventional color gradient is obtained by fitting a straight line to 
the color profile, which neglects any curvature that may be present. Also,
due to the limited signal-to-noise available, often the errors
on the color gradient can be large, making the measured values
uncertain. 
The process of obtaining
scale lengths involve averaging over isophotes as well as the profile fit
with an empirically tested model.
We have seen above that the $\chisqr$ obtained are well within
acceptable limits in most cases for good fits.  The scale lengths are
therefore good, robust indicators of the large scale distribution of
light in the galaxy, and their ratio in the two filters provides a useful
descriptor of the way the color changes over the galaxy.  
Using the ratio we have demonstrated the ubiquity of inner blue color in 
radio galaxies, a fact which is not apparent from the conventional color gradient. 

\subsection{Discussion}

Color gradients in early type galaxies are believed to be due to 
metallicity and age gradients in the stellar population.  The presence of
dust produces increased reddening, while star formation leads to bluer colors.
In addition to any overall color gradient that we observe in the radio galaxies,
their $B-R$ color images show regions of excess reddening indicative of dust.
The dust occurs in the form of coherent lanes in  $\sim20\%$ of the galaxies,
while another $\sim17\%$ show dust patches.  The remaining objects could of course contain dust well mixed with stars, which is not evident in the color maps.  In several of the galaxies with $\reb/\rer>1$, we see clear 
clear evidence of dust in the color maps.  
In the control galaxies detectable  dust again occurs in $\sim37\%$
of the sample, but here the dust is more often patchy ($30\%$). 

Assuming that the
composition of dust in the radio sample is similar to that in our Galaxy, 
and by using a simple screen model with constant gas-to-dust ratio 
\REFR (see \eg Burstein \& Heiles 1978),
the mass of the dust can be estimated in the usual manner from the
excess $B-R$ color.  The dust mass turns out to be in the range
$\sim10^5-10^7\msol$.   A plot of dust mass against radio power 
(see \fig{fig4})
shows that the two are correlated,  the linear correlation coefficient  
being significant at better than the $99\%$ confidence level.  The dust mass
as well as the luminosity depend on the square of the distance, which 
could lead to a false correlation.  The distance effect is probably not 
serious in the present case since the partial correlation coefficient,
obtained after factoring out the distance, remains significant at better 
than the $90\%$ level.

We have seen above that the  distribution of the $r_e(B)/r_e(R)$ in the
radio galaxies indicates that these objects more often become bluer towards the
center than the control galaxies.  The blue color is presumably due
to star formation which is in some manner induced by the presence of the
radio source.  We have shown in \fig{fig4} a plot
of the logarithm of the total radio power at $408\mhz$ against 
$\log(r_e(B)/r_e(R))$.  It is seen that there is a clear trend for the more 
powerful radio galaxies to have steeper color gradients as indicated by the 
scale length ratio.   The more luminous a radio source is, the greater seems 
to be the increased blue luminosity triggered by it.  
\REFR Best, Longair, \& R\"{o}ttgering (1996) report finding either 
a string of bright star forming knots
or compact knots in the case of $z \sim1$ radio galaxies from the 3CR catalogue.
These are thought to be produced by the interaction of the radio jet 
with the interstellar medium.  It will be possible to model the
mass and spatial extent of the gas involved in the bursts from narrow band
imaging and long-slit spectra of the galaxies.    

\section{Conclusions}

Using detailed model profile fits to the observed
surface brightness profiles of samples of radio and control galaxies,
we have shown that  the distribution of the $r_e(B)/r_e(R)$ ratio
is different for the two samples.  A value $<1$ for this ratio
in a galaxy indicates that the color of the galaxy becomes bluer towards the
center, while $r_e(B)/r_e(R)>1$ indicates that the color becomes redder
towards the center.  The ratio has a simple relation with  the
color gradient $G$ obtained directly from color profiles.  But the
ratio can be used as an indicator of the radial dependence of
color, even when the measured gradient is too noisy to serve the purpose.

{}

\eject

\end{document}